\documentclass[cmp]{svjourmod}

\usepackage{dcolumn}% Align table columns on decimal point
\usepackage{bm}% bold math
\usepackage{verbatim}       % Defines \begin{comment}, \end{comment}

\usepackage[dvips]{graphicx}
\usepackage{amssymb}
\usepackage{bm}
\usepackage{amsmath}
\usepackage{amsfonts}
\newcommand{\dd}{{\rm d}}
\begin{document}
%
%\DeclareGraphicsExtensions{.pdf}

\title{Time functions as utilities}

%\begin{comment}
%\author{E. Minguzzi \footnote{Dipartimento di Matematica Applicata, Universit\`a degli Studi di Firenze,  Via
%S. Marta 3,  I-50139 Firenze, Italy. E-mail:
%ettore.minguzzi@unifi.it}}
%\end{comment}

%\author{E. Minguzzi\thanks{
%Dipartimento di Matematica Applicata ``G. Sansone'', Universit\`a
%degli Studi di Firenze, Via S. Marta 3,  I-50139 Firenze, Italy.
%E-mail: ettore.minguzzi@unifi.it} }

\author{E. Minguzzi}
\institute{Dipartimento di Matematica Applicata, Universit\`a degli
Studi di Firenze,  Via S. Marta 3,  I-50139 Firenze, Italy \\
\email{ettore.minguzzi@unifi.it}
%\\
%Phone: +39 055 4796 253, Fax: +39 055 471787
}
\authorrunning{E. Minguzzi}

\date{}
\maketitle

\date{}
\maketitle

\begin{abstract}
\noindent Every time function on spacetime gives a (continuous)
total preordering of the spacetime events which respects the notion
of causal precedence. The problem of the existence of a (semi-)time
function on spacetime and the problem of recovering the causal
structure starting from the set of time functions are studied. It is
pointed out that these problems have an analog in the field of
microeconomics known as utility theory. In a chronological spacetime
the semi-time functions correspond to the utilities for the
chronological relation, while in a
 $K$-causal (stably causal) spacetime the time functions correspond
to the utilities for the $K^{+}$ relation (Seifert's relation). By
exploiting this analogy, we are able to import some mathematical
results, most notably Peleg's and Levin's theorems, to the spacetime
framework. As a consequence, we prove that a $K$-causal (i.e. stably
causal) spacetime admits a time function and that the time or
temporal functions can be used to recover the $K^{+}$ (or Seifert)
relation which  indeed turns out to be the intersection of the time
or temporal orderings. This result tells us in which circumstances
it is possible to recover the chronological or causal relation
starting from the set of time or temporal functions allowed by the
spacetime. Moreover, it is proved that a chronological spacetime  in
which the closure of the causal relation  is transitive (for
instance a reflective spacetime) admits a semi-time function. Along
the way a new proof avoiding smoothing techniques is given that the
existence of a time function implies stable causality, and a new
short proof of the equivalence between $K$-causality and stable
causality is given which takes advantage of Levin's theorem and
smoothing techniques.

%
%Every time function on spacetime gives a (continuous) total ordering
%of the spacetime events which respects the notion of causal
%precedence. It is proved here that in a stably causal spacetime the
%intersection of all the so obtained total orderings gives  Seifert's
%partial order. This result tells us in which circumstances it is
%possible to recover the chronological or causal relation starting
%from the set of time or temporal functions allowed by the spacetime.
%Moreover,  a new proof avoiding smoothing techniques is given that
%the existence of a time function implies stable causality. The
%results of this paper are consequence of the recently proved
%equivalence between $K$-causality and stable causality.
\end{abstract}

%\pacs{}

%\noindent Key Words:

\section{Introduction}

%Indeed, given a spacetime
%$(M,g)$ the causal structure is identified with the class $\bf{g}$
%of metrics conformal to $g$, as this class alone determines  the
%light cones on spacetime.
On the spacetime $(M,g)$ we write as usual $p<q$ if there is a
future directed causal curve connecting $p$ to $q$, and write $p\le
q$ if $p<q$ or $p=q$. The causal relation is given by
$J^{+}=\{(p,q)\in M\times M: p\le q\}$. For fixed time orientation
these notions depend only on the the class $\bf{g}$ of metrics
conformal to $g$. Once the causal relation is defined it is possible
to define a {\em time function} as a continuous function $t: M \to
\mathbb{R}$ such that if $p<q$ then $t(p)<t(q)$. In other words a
time function is defined all over the spacetime, it is continuous,
and increases over every causal curve. A time function which is
$C^1$ with a past directed timelike gradient is a {\em temporal}
function. Following \cite{seifert77}, a semi-time function is a
continuous function $t: M \to \mathbb{R}$ such that if $p\ll q$ then
$t(p)<t(q)$ (note that by continuity we have also $(p,q) \in
\overline{I^{+}} \Rightarrow t(p)\le t(q)$).

These definitions clarify that in the framework of general
relativity the notion of causality is more fundamental than that of
time. Indeed, not all the spacetimes admit a time function. A
spacetime admits a time function iff it admits a temporal function
iff  it is stably causal \cite{hawking68,hawking73,bernal04}. The
history of this result is quite interesting.

In order to prove the existence of a time function Geroch
\cite{geroch70} suggested to introduce a positive measure $\mu$ on
spacetime so that $M$ has unit measure (in fact this measure has to
be chosen so as to satisfy some admissibility constraints
\cite{dieckmann88}), and to define $t^{-}(p)=\mu(I^{-}(p))$. In
globally hyperbolic spacetimes, and actually in causally continuous
ones \cite{hawking74,dieckmann88,minguzzi06c}, the idea works in
fact $t^{-}$ can be shown to be continuous. Nevertheless, these
causality conditions are stronger than stable causality and without
them $t^{-}$ is only lower semi-continuous.

The proof that stable causality implies the existence of a time
function was obtained by Hawking through a nice averaging technique
\cite{hawking68,hawking73}. In short he noted that if the spacetime
is stably causal then there is a one parameter family of metrics
with cones strictly larger than $g$, $g_\lambda>g$ with $\lambda\in
[1,2]$, $\lambda<\lambda' \Rightarrow g_\lambda<g_{\lambda'}$, so
that $(M,g_\lambda)$ is causal. He then defined $t(p)=\int_1^2
t^{-}_\lambda(p)\,\dd \lambda$ where the function $t^{-}_\lambda$ is
defined as before but with respect to the metric $g_\lambda$. He was
then able to prove the continuity of $t$ (see \cite{hawking73}).

The proof of the converse presents several difficulties particularly
because a time function $t$ for $(M,g)$ need not be  a time function
for some $(M,g')$ with $g'>g$, consider for instance $t=x^0-\tanh
x^1$ in the 1+1 Minkowski spacetime of metric $g=-(\dd x^0)^2+(\dd
x^1)^2$. Nevertheless, the proof that a temporal function implies
stable causality is easy \cite{hawking73} and thus there remained
the issue of proving that the existence of a time function implies
the existence of a temporal function. This smoothability problem was
considered by Seifert \cite{seifert77} but his arguments were
unclear. A rigorous proof was finally given by Bernal and S\'anchez
in \cite{bernal04}.

%This however is only part of the story, further insights come from
%the relational approach.

Further insight to the problem of the existence of time come from
the relational approach to causality. Stable causality can be shown
to be equivalent to the antisymmetry of the Seifert relation
\cite{seifert71} $J^{+}_S=\bigcap_{g'>g}J^{+}_{g'}$ (a rigorous
proof can be found in \cite{hawking74} and \cite{minguzzi07}). The
nice feature of this relation is that it is both closed and
transitive whereas $J^{+}$ has only the latter property and
$\overline{J^{+}}$ has only the former. In fact one may ask if
$J^{+}_S$ is the smallest relation containing $J^{+}$ with this
property. The answer is negative unless some causality conditions
are added \cite{minguzzi07}. Therefore, it is natural to introduce
the relation $K^{+}$ defined
 as the smallest closed and transitive relation which
contains $J^{+}$ (see \cite{sorkin96}). The spacetime is said to be
$K$-causal if $K^{+}$ is antisymmetric. Recently \cite{minguzzi08b},
I have proved the equivalence between $K$-causality and stable
causality and that if $K$-causality holds then $K^{+}=J^{+}_S$.

The previous result shows that the antisymmetry of $K^{+}$ implies
stable causality and hence the existence of a time function. It
seems reasonable to expect that (i): this theorem depends only on
the transitivity and closure properties of $K^{+}$, and that
therefore passing through stable causality should not be  essential.
(ii): The existence of a time function should imply $K$-causality
(or stable causality) directly without using the smoothing argument.
Finally, given the fact that $K$-causality implies the existence of
a time function one would like to prove that (iii): under stable
causality the set of time functions allowed by the spacetime can be
used to recover the relation $K^{+}$.

In a first version of this work I presented proofs for points (ii)
and (iii) but then searching for fundamental results using only the
closure of a relation in connection with problem (i), I discovered a
large body of literature in utility theory with important
implications for causality (most articles were published in
economics journals). In fact the problem of the existence of a
utility function in a set of alternatives for an individual is
formally similar to that of the existence of time. Surprisingly,
these results have been totally overlooked by relativists. In the
next section I summarize this long parallel line of research which
will be used to draw implications for causality theory and in
particular for the problem of the existence of time.

I refer the reader to \cite{minguzzi06c,minguzzi07b} for most of the
conventions used in this work. In particular, I denote with $(M,g)$
a $C^{r}$ spacetime (connected, time-oriented Lorentzian manifold),
$r\in \{3, \dots, \infty\}$ of arbitrary dimension $n\geq 2$ and
signature $(-,+,\dots,+)$. On $M\times M$ the usual product topology
is defined.  All the causal curves that we shall consider are future
directed. The subset symbol $\subset$ is reflexive, $X \subset X$.

%
%
%In this work I shall give a  direct relational proof that the
%existence of a time function implies stable causality. The new proof
%does not use smoothing techniques, thus it does not pass through the
%existence of a temporal function. Instead, it uses the recently
%proved equivalence between $K$-causality and stable causality
%\cite{minguzzi08b}. In this respect it is useful to recall that
%$K^{+}$ is defined \cite{sorkin96} as the smallest closed and
%transitive relation which contains $J^{+}$. The spacetime is said to
%be $K$-causal if $K^{+}$ is antisymmetric. Since smoothing
%techniques are not used and the new proof is relational in nature,
%it can perhaps be generalized to more abstract settings in which the
%metric is only $C^0$ \cite{sorkin96}  or  the causal relations are
%given without reference to a Lorentzian metric as in
%\cite{kronheimer67}.

\section{Preorders and utility theory} \label{secb}

Recall\footnote{Unfortunately, in the literature there is no
homogeneous terminology, so that used in this paper differs from
that of many cited articles.} that a binary relation $R\subset
X\times X$ on a set $X$
   is called a  {\em  preorder} if it is reflexive and
transitive, a {\em strict partial order} if it is irreflexive and
transitive, and an {\em equivalence relation} if it is reflexive,
transitive and symmetric. A preorder which satisfies the
antisymmetry property $(x,y)\in R$ and $(y,x) \in R \Rightarrow
x=y$,  is a {\em partial order}. The preorder or strict partial
order $R$ such that $x\ne y \Rightarrow (x,y)\in R$ or $(y,x) \in
R$,
 is {\em complete}. The property, if $a,b \in X$ then
$(a,b)\in R$ or $(b,a)\in R$ is the {\em totality} property and is
equivalent to completeness and reflexivity. A preorder which
satisfies the totality property, is a {\em  total preorder}. A
preorder which respects both the totality and the antisymmetry
property is a  {\em total order}. A strict partial order which is
complete is a {\em complete order}.

For short we also write\footnote{Unfortunately, if $R=J^{+}$ then
while $\le_{J^{+}}$ has the same meaning of the symbol $\le$ in
relativity, the relation $<_{J^{+}}$ coincides with $<$ only in
causal spacetimes.} $x \le_R y$ if $(x,y) \in R$; $x\sim_R y$ if
$(x,y)\in R$ and $(y,x)\in R$; and $x<_R y$ if $x \le_R y$ and not
$x\sim_R y$. The relation $\sim_R$ is an equivalence relation called
the {\em equivalence relation part} while $<_R$ is a strict partial
order called the {\em strict partial order part}. Their union gives
$R$, i.e. $x\le_R y$ iff $x\sim_R y$ or $x <_R y$. Note that if $R$
is a partial order then $<_R$ is obtained from $R$ by removing the
diagonal $\Delta=\{(x,x): x \in X \times X\}$, and conversely $R$ is
obtained by adding the diagonal to $<_R$, that is, $R$ is the
smallest reflexive relation containing $<_R$ (the reflexive
closure).

 Given a (total) preorder $R$,   the quotient $X/\sim_R$  endowed
with the induced order $[p]\le_R [q]$ iff $p\le_R q$, is a partial
(resp. total) order.

As usual we denote $R^{+}(x)=\{y\in X:(x,y) \in R\}$ and
$R^{-}(x)=\{y\in X:(y,x) \in R\}$.

We say that $R_2$ extends $R_1$ if $x\sim_{R_1} y \Rightarrow
x\sim_{R_2} y$ and $x<_{R_1} y \Rightarrow x<_{R_2} y$.
%
%A strict partial order $S$ is  irreflexive, i.e. $(p,p)\notin S$,
%and transitive binary relation, thus in particular it is asymmetric:
%$(c,d)\in S \Rightarrow (d,c) \notin S$. A strict total order is a
%strict partial order which is also trichotomic: $(a,b) \in S$ or
%$a=b$ or $(b,a) \in S$.
%
%There is a one-to-one correspondence between partial orders and
%strict partial orders obtained by adding the diagonal of the ambient
%space $\Delta=\{(x,x): x \in X \times X\}$, $S=R\backslash \Delta$,
%$R=S\cup \Delta$.

%Edward Marczewski (his surname until 1940 was Szpilrajn)
Through a rather simple application of Zorn's lemma Szpilrajn
\cite{szpilrajn30} proved

\begin{theorem} \label{szp}
(Szpilrajn) Every strict partial order can be extended to a complete
order. Moreover, every strict partial order is the intersection of
all the complete orders that extend it. (by adding the diagonal one
has a corresponding statement for partial orders extended by total
orders.)
\end{theorem}
The former statement in the theorem is also known as the {\em order
extension principle}. The latter statement is sometimes attributed
to Dushnik and Miller \cite{dushnik41}, however, although not stated
explicitly in \cite{szpilrajn30}, it is a trivial consequence of a
remark in Szpilrajn's paper.

Since to any preorder one can associate a  partial order passing to
the quotient with respect to the equivalence relation $\sim_R$, it
is easy  to prove from theorem \ref{szp} the following
\cite{donaldson98,bossert99}

\begin{theorem}
Every preorder can be extended to a total preorder. Moreover, every
preorder is the intersection of all the total preorders that extend
it.
\end{theorem}
I note that the proof in \cite{bossert99} is such that the total
extension can be chosen (or restricted in the second part) in such a
way that the `indifference sets' $[p]$ are not enlarged passing from
the preorder $R$ to its total extension $C$, i.e. $x\sim_{R} y
\Leftrightarrow x\sim_{C} y$.

These results were taken as reference for many other developments,
and in fact have been generalized in several directions
\cite{andrikopoulos09}.

Meanwhile, in microeconomics the preference of an individual for a
set of alternatives or prospects $X$ was modeled as a {\em total
preorder} $R$ on $X$. The idea was that an individual is able to
tell whether one option or the other is preferred. These preferences
were quantified by an {\em utility function}. An utility for a
transitive relation $R$ is a function $u: X \to \mathbb{R}$ with the
{\em strictly isotone}\footnote{A function is isotone if $(x,y)\in R
\Rightarrow u(x)\le u(y)$. Note that constant functions are
isotone.} property (citare birkoff) namely that\footnote{For a total
preorder this definition can be replaced by: $x \le_R y$ if and only
if $u(x)\le u(y)$.}
\begin{equation} \label{uti}
''x \sim_R y \Rightarrow u(x)= u(y)'' \ \textrm{ and } \ ''x <_R y
\Rightarrow u(x)< u(y).''
\end{equation}
Often on $X$ one has a topology which makes rigorous the idea that
an alternative is similar or close to another. In this case one
would like to have a {\em continuous} utility function otherwise the
closeness of the alternatives would not be correctly represented by
the utility. Eilenberg \cite{eilenberg41} and Debreu
\cite{debreu54,debreu64} (see also \cite{rader63,mehta77}) were able
to prove, under weak topological assumptions,  that a continuous
utility exists provided $R^{-}(x)$ and $R^{+}(x)$ are closed for
every $x \in X$.

With the work of Aumann \cite{aumann62} and other economists it
became clear that the assumption of totality was too restrictive. It
turns out that it is unreasonable to assume that the individual is
able to decide the preference for one of two alternatives. This
conclusion is even more compelling if one models a group of persons
rather than an individual. In order to include some indecisiveness
the space of alternatives $X$ has to be endowed with a preorder, the
totality condition being removed. The previous results on the
existence of a continuous utility must therefore be generalized, and
there is indeed a large literature on the subject. The reader is
referred to the monograph by Bridges and Mehta for a nice reasoned
account \cite{bridges95}.

The problem is of course that of finding natural conditions which
imply the continuity of the utility function. We already suspect,
given the suggestions from the spacetime problem above, that this
property is the closure of the relation $R$ in $X\times X$,
$R=\bar{R}$ (sometimes called {\em continuity} in the economics
literature). In fact, it can be easily shown \cite{ward54} that for
a total preorder this property coincides with that used in Eilenberg
and Debreu theorem, namely: $R^{-}(x)$ and $R^{+}(x)$ are closed for
every $x \in X$ (sometimes called {\em semicontinuity} in the
economics literature).

In the literature many other direction have been explored but as we
shall see the closure condition  has given the most powerful
results. Considerably important has proved the work by Nachbin
\cite{nachbin65}, who  studied in deep closed preorders on
topological spaces and obtained an extension to this domain of the
Urysohn separation and extension theorems. These results were used
to obtain new proofs of the Debreu theorem \cite{mehta77}, and over
them have been built the subsequent generalizations. For our
pourposes, the final goal was reached
 by Levin \cite{levin83,bridges95} who proved
\begin{theorem} \label{lev} (Levin)
Let $X$ be a second countable locally compact Hausdorff space, and
$R$ a closed preorder on $X$, then there exists a continuous utility
function. Moreover, denoting with $\mathcal{U}$  the set of
continuous utilities
 we have that the
preorder $R$ can be recovered from the continuous utility functions,
namely there is a {\em multi-utility representation}
\begin{equation}
(x,y) \in R \Leftrightarrow \forall u \in \mathcal{U}, \ u(x)\le
u(y).
\end{equation}
\end{theorem}
Curiously, as it happened for Szpilrajn's theorem, the second part
of this statement was not explicitly given in Levin's paper.
Nevertheless, it is a trivial consequence of his proof that if
$x\nleq_R y$ then there is a continuous utility function such that
$u(x)>u(y)$ (see the end of the proof of \cite[Lemma
8.3.4]{bridges95}). Apparently, this representation possibility has
been pointed out only quite recently in a  preprint by Evren and Ok
\cite{evren09}.

Clearly, Levin's theorem can be regarded as the continuous analog of
the Szpilrajn's theorem.

It is curious to note that while there was, as far as I know, no
communication between the communities of relativists and economists
these two parallel lines of research passed through the very same
concepts. For instance, Sondermann \cite{sondermann80} (see also
\cite{candeal93}) introduced a measure on $X$ and built an
increasing  function exactly as Geroch did, even more his
admissibility requirements for the measure were close to those later
introduced by Dieckmann \cite{dieckmann88} in the relativity
literature. As expected he could only obtain lower semi-continuity
for the utility, indeed we know that Geroch's time is continuous
only if some form of reflectivity is imposed \cite{hawking74}.

Among the theorems which do not use in an essential way some closure
assumption, which in most cases can be deduced from Levin's theorem,
a special mention deserves Peleg's theorem which instead uses an
openness condition \cite{peleg70}. According to Peleg a {\em strict}
partial order $S$ is {\em separable} if there is a countable subset
$C$ of $X$ such that for any  $(x,y) \in S$ the diamond
$S^{+}(x)\cap S^{-}(y)$ contains some element of the subset $C$, and
{\em spacious} if for $(x,y)\in S$, $\overline{S^{-}(x)} \subset
S^{-}(y)$.
\begin{theorem} (Peleg)
Let $S$ be a strict partial order on a topological space $X$.
Suppose that (a) $S^{-}(x)$ is open for every $x \in X$, (b) $S$ is
separable, and (c) $S$ is  spacious, then there is a function $u:
X\to \mathbb{R}$ such that $(x,y)\in S \Rightarrow u(x)<u(y)$.
\end{theorem}
Note that $u$ is an utility in the sense of Eq. (\ref{uti}). It has
been clarified that Debreu's theorem can be regarded as a
consequence of Peleg's \cite{lee72,mehta83}, while the relation with
Levin's theorem is less clear (but see \cite{herden89,herden95}). We
will be able to say more on that in the next section where  we shall
apply the previous theorems to the spacetime case.

\section{Application of utility theory to causality}

Our strategy will be that of applying Peleg's theorem to the open
relation $I^{+}$ and Levin's theorem to the closed relation $K^{+}$.
As we shall see the semi-time functions and the time functions
correspond to the utilities for the relations $I^{+}$ and $K^{+}$
respectively, provided they are antisymmetric.

We start with the former case

\subsection{Semi-time functions and $I^{+}$-utilities}

In this section we apply Peleg theorem by setting $X=M$, the
spacetime manifold, and $S=I^{+}$. The assumption that $I^{+}$ is a
strict partial order is equivalent to chronology and conditions (a)
and (b) of Peleg's theorem are satisfied. Note that in a
chronological spacetime a  continuous utility for $I^{+}$ is  a
continuous function $t$ such that $x \ll y \Rightarrow t(x)<t(y)$,
thus the continuous utilities for $I^{+}$ are exactly the semi-time
functions. Therefore we have only to understand the condition of
spaciousness: $x \ll y \Rightarrow \overline{I^{-}(x)} \subset
I^{-}(y)$. This problem is answered by the following

\begin{lemma}
Let $(M,g)$ be a spacetime, then the spaciousness condition, $x \ll
y \Rightarrow \overline{I^{-}(x)} \subset I^{-}(y)$, is equivalent
to the future reflectivity condition, $p \in \overline{I^{-}(q)}
\Rightarrow q \in \overline{I^{+}(p)}$.
\end{lemma}
%\begin{remark}
(The previous definition of future reflectivity is equivalent to the
usual one $I^{-}(w)\subset I^{-}(z)\Rightarrow I^{+}(z)\subset
I^{+}(w)$, see \cite{hawking74,minguzzi06c}.)
%\end{remark}

\begin{proof}
Assume that $(M,g)$ is future reflective, take $x \ll y$ and let
$z\in \overline{I^{-}(x)}$ then $x \in \overline{I^{+}(z)}$ and
since $I^{-}(y)$ is open, $z\ll y$. As $z$ is arbitrary
$\overline{I^{-}(x)}\subset I^{-}(y)$.

Conversely, assume the spacetime is spacious. Let $p \in
\overline{I^{-}(q)}$ and take $r \in I^{+}(q)$ then by spaciousness,
$I^-(r)\supset \overline{I^{-}(q)}\ni p$ so that $r \in I^{+}(p)$.
As $r$ can be chosen arbitrarily close to $q$, we have $q \in
\overline{I^{+}(p)}$. $\square$
\end{proof}

With these preliminaries, Peleg's theorem becomes (in the statement
we include the past version)

\begin{theorem} \label{weak}
A chronological, future or past reflective spacetime admits a
semi-time function.
\end{theorem}

This theorem is  new in causality theory as there were no previous
results establishing the existence of a semi-time function. We
observe that if the causality assumption were somewhat stronger,
with chronology replaced by  distinction ($I^{+}(x)=I^{+}(y)$ or
$I^{-}(x)=I^{-}(y) \Rightarrow x=y$) then the spacetime would be
$K$-causal \cite[Theorem 3.7]{minguzzi07b}, thus it would admit a
time function.

Also note that if future or past reflectivity is strengthened to
reflectivity then the chronological assumption can be weakened to
non-total viciousness (namely the chronology violating set does not
coincide with $M$). This fact follows from a theorem by Clarke and
Joshi which states that a non-totally vicious spacetime which is
reflective is chronological \cite[Prop. 2.5]{clarke88} (see also
\cite{kim93}).

In \cite{minguzzi08b} I have introduced the {\em transverse
conformal ladder} and proved that future or past reflectivity
implies the transitivity of $\overline{J^{+}}$, that is
$K^{+}=\overline{J^{+}}$, (see the proof of theorem 3 in
\cite{minguzzi08b}). Thus one could try to strengthen the previous
theorem by replacing `future or past reflectivity' with the
transitivity of $\overline{J^{+}}$. As we shall see in the next
section it is indeed possible to do that.

\subsection{Time functions and $K^{+}$-utilities}
Any spacetime is a paracompact Hausdorff manifold and as such it
satisfies the topological conditions of Levin's theorem (in fact it
even admits a complete Riemannian metric \cite{nomizu61}). Since we
wish to apply this theorem to the relation $K^{+}$ we have first to
establish the relation between the time functions and the continuous
$K^{+}$-utilities.

In this section we shall prove, without using smoothing techniques
\cite{bernal04} or the equivalence between $K$-causality and stable
causality \cite{minguzzi08b}, that a spacetime is $K$-causal if and
only if it admits a time function. Along the way we shall also prove
that in a $K$-causal spacetime the $K^{+}$-utilities are exactly the
time functions.

\begin{lemma} \label{jod}
A spacetime which admits a time function $t$  is strongly causal.
\end{lemma}

\begin{proof}
The proof of \cite[Theorem 3.4]{minguzzi07b} shows that if $(M,g)$
is not strongly causal then there are points $x,c \in M$ such that
$x<c$ and $(c,x) \in \overline{J^{+}}$. Since for any pair $(p,q)\in
J^{+}$ it is $t(q)-t(p)\ge 0$ and $t$ is continuous, $t(x)<t(c)\le
t(x)$, a contradiction. $\square$
\end{proof}

Recall that a spacetime is non-total imprisoning if no future
inextendible causal curve is contained in a compact set (replacing
{\em future} with {\em past} gives the same property
\cite{beem76,minguzzi07f}). Strong causality implies non-total
imprisonment \cite{hawking73}.

\begin{lemma} \label{pkw}
Let $(M,g)$ be non-total imprisoning. Let $(p,q)\in K^{+}$ then
either $(p,q)\in J^{+}$ or for every relatively compact open set
$B\ni p$ there is $r \in \dot{B}$ such that $p<r$ and $(r,q)\in
K^{+}$.
\end{lemma}

\begin{proof}
Consider the relation
\begin{align*}
R^{+}=\{(p,q)\in K^{+}: &\ (p,q)\in J^{+} \textrm{ or for every
relatively compact open set } B\ni p \\
&  \textrm{ there is } r \in \dot{B} \textrm{ such that } p<r
\textrm{ and } (r,q)\in K^{+} \}.
\end{align*}
It is easy to check that $J^{+}\subset R^{+}\subset K^{+}$. We are
going to prove that $R^{+}$ is closed and transitive. From that and
from the minimality of $K^{+}$ it follows $R^{+}=K^{+}$ and hence
the thesis.

Transitivity: assume $(p,q) \in R^{+}$ and $(q,s) \in R^{+}$. If
$(p,s) \in J^{+}$ there is nothing to prove.  Otherwise we have
$(p,q)\notin J^{+}$ or $(q,s) \notin J^{+}$. Let $B\ni p$ be an open
relatively compact set.

If $(p,q)\notin J^{+}$  there is $r \in \dot{B}$ such that $p<r$ and
$(r,q)\in K^{+}$, thus $(r,s) \in K^{+}$ and hence $(p,s) \in
R^{+}$.

It remains to consider the case $(p,q)\in J^{+}$ and $(q,s) \notin
J^{+}$. If $p=q$ then $(p,s)=(q,s) \in R^{+}$. Otherwise, $p<q$ and
if $q \notin B$ the causal curve $\gamma$ joining $p$ to $q$
intersects $\dot{B}$ at a point $r \in \dot{B}$ (possibly coincident
with $q$ but different from $p$). Thus $p<r$, $(r,q) \in J^{+}$,
hence $p<r$ and $(r,s) \in K^{+}$. If $q \in B$, since $(q,s)\in
R^{+}\backslash J^{+}$, there is $r \in \dot{B}$ such that $q<r$ and
$(r,s) \in K^{+}$, moreover, since $p\le q$, $p<r$. Since the
searched conclusions holds for every $B$, $(p,s) \in R^{+}$.

Closure: let $(p_n,q_n) \to (p,q)$, $(p_n,q_n) \in R^{+}$. Assume,
by contradiction, that $(p,q) \notin R^{+}$, then $p\ne q$ as
$J^{+}\subset R^{+}$. Without loss of generality we can assume two
cases: (a) $(p_n,q_n) \in J^{+}$ for all $n$; (b) $(p_n,q_n) \notin
J^{+}$  for all $n$.

(a) Let $B\ni p$ be an open relatively compact set. For sufficiently
large $n$, $p_n \ne q_n$ and $p_n \in B$. By the limit curve theorem
\cite{minguzzi07c} either there is a limit continuous causal curve
joining $p$ to $q$, and thus $p<q$ (a contradiction), or there is a
future inextendible continuous causal curve $\sigma^p$ starting from
$p$ such that for every $p'\in \sigma^p$, $(p',q) \in
\overline{J^{+}}$. Since $(M,g)$ is non-total imprisoning,
$\sigma^p$ intersects $\dot{B}$ at some point $r$. Thus $p<r$ and
since $(r,q)\in \overline{J^{+}}\subset K^{+}$ we have $(p,q)\in
R^{+}$, a contradiction.

(b) Let $B\ni p$ be an open relatively compact set. For sufficiently
large $n$, $p_n \ne q_n$ and $p_n \in B$. Since $(p_n,q_n) \in
R^{+}$ there is $r_n \in \dot{B}$, $p_n<r_n$, and $(r_n,q_n)\in
K^{+}$. Without loss of generality we can assume $r_n \to r \in
\dot{B}$, so that $(r,q) \in K^{+}$. Arguing as in (a) either $p<r$
(and $(r,q)\in K^{+}$) or there is $r'\in \dot{B}$ such that $p<r'$
and $(r',r) \in \overline{J^{+}}\subset K^{+}$, from which is
follows $(r',q) \in K^{+}$. Because of the arbitrariness of $B$,
$(p,q)\in R^{+}$, a contradiction. $\square$

\end{proof}

\begin{corollary}
Let $(M,g)$ be non-total imprisoning. The spacetime $(M,g)$ is
$K$-causal if and only if $x<y$ and $(y,x)\in K^{+}$ implies $x=y$.
\end{corollary}

\begin{proof}
Since $J^{+}\subset K^{+}$ to the right it is trivial. To the left,
assume $(M,g)$ is not $K$-causal, then there are $z,y \in M$, $z \ne
y$ such that $(z,y)\in K^{+}$ and $(y,z)\in K^{+}$. If $(z,y)\in
J^{+}$ we have finished with $x=z$. Otherwise, let $B\ni z$ be an
open relatively compact set. By lemma \ref{pkw} there is a point
$x\in \dot{B}$ such that $z<x$ and $(x,y)\in K^{+}$ (and thus $(x,z)
\in K^{+}$) which implies $z=x \in \dot{B}$, a contradiction.
$\square$
\end{proof}

\begin{lemma} \label{jba}
$\empty$
\begin{description}
\item[(a)] Let $\tilde{t}$ be a continuous function such that $x\le y \Rightarrow
\tilde{t}(x)\le \tilde{t}(y)$. If $(p,q) \in K^{+}$ then
$\tilde{t}(p)\le \tilde{t}(q)$.
\item[(b)] Let $t$ be a time function on $(M,g)$. If
$(p,q) \in K^{+}$ then  $p=q$ or $t(p)<t(q)$.
\end{description}
\end{lemma}

\begin{proof}

Proof of (a). Consider the relation
\[
\tilde{R}^{+}=\{(p,q)\in K^{+}:  \ \tilde{t}(p)\le \tilde{t}(q) \}.
\]
Clearly $J^{+}\subset \tilde{R}^+\subset K^{+}$ and $\tilde{R}^+$ is
transitive.

 %Consider the relation
%\begin{align*}
%\tilde{R}^{+}&=\{(p,q)\in K^{+}:  \ \tilde{t}(p)\le \tilde{t}(q) \}, \\
%R^{+}&=\{(p,q)\in K^{+}: \ p=q \textrm{ or } t(p)<t(q) \} .
%\end{align*}
%Clearly, they both contain $J^{+}$ and are contained in  $K^{+}$. It
%is also trivial to check that they are both transitive.

Let us prove that $\tilde{R}^{+}$ is closed. If $(x_n,z_n) \in
\tilde{R}^{+}$ is a sequence such that $(x_n,z_n) \to (x,z)$, then
passing to the limit $\tilde{t}(x_n)\le \tilde{t}(z_n)$ and using
the continuity of $\tilde{t}$ we get $\tilde{t}(x)\le \tilde{t}(z)$,
moreover since $K^{+}$ is closed, $(x,z) \in K^{+}$, which implies
$(x,z) \in \tilde{R}^{+}$, that is $\tilde{R}^{+}$ is closed.

Since $J^{+}\subset \tilde{R}^{+}\subset K^{+}$, and $\tilde{R}^{+}$
is closed and transitive, by using the minimality of $K^{+}$ it
follows that $\tilde{R}^{+}=K^{+}$. As a consequence, if $(p,q)\in
K^{+}$ then $\tilde{t}(p)\le \tilde{t}(q)$.

Proof of (b).  By lemma \ref{jod}, since $t$ is a time function
$(M,g)$ is strongly causal and thus non-total imprisoning. Consider
the relation
\[
R^{+}=\{(p,q)\in K^{+}: \ p=q \textrm{ or } t(p)<t(q) \} .
\]
Clearly $J^{+}\subset {R}^+\subset K^{+}$ and ${R}^+$ is transitive.
Let us prove that $R^{+}$ is closed by keeping in mind the just
obtained result given by (a).  Let $(p_n,q_n) \in {R}^{+}\subset
K^{+}$ be a sequence such that $(p_n,q_n) \to (p,q)$. As $K^{+}$ is
closed, $(p,q) \in K^{+}$. If, by contradiction, $(p,q) \notin
R^{+}$ then $(p,q) \notin J^{+}$,  thus by lemma \ref{pkw}, chosen
an open relatively compact set $B\ni p$ there is $r\in \dot{B}$,
with $p<r$, $(r,q)\in K^{+}$, thus $t(p) <t(r)\le t(q)$ and hence
$(p,q)\in R^{+}$, a contradiction.

Since $J^{+}\subset {R}^{+}\subset K^{+}$, and ${R}^{+}$ is closed
and transitive, by using the minimality of $K^{+}$ it follows that
${R}^{+}=K^{+}$. As a consequence, if $(p,q)\in K^{+}$ then either
$p=q$ or $t(p)< t(q)$. $\square$

\end{proof}

\begin{theorem} \label{ndc}
In a $K$-causal spacetime the continuous $K^{+}$-utilities are
 the time functions.
\end{theorem}

\begin{proof}
A $K^{+}$-utility is a function $u$ which satisfies (i) $(x,y)\in
K^{+}\Rightarrow u(x)\le u(y)$ and (ii) $(x,y)\in K^{+}$ and $(y,x)
\notin K^{+}\Rightarrow u(x)<u(y)$. Since the spacetime is
$K$-causal this condition is equivalent to $(x,y)\in K^{+}
\Rightarrow x=y$ or $u(x)<u(y)$. Thus by lemma \ref{jba} point (b),
every time function is a continuous $K^{+}$-utility. Conversely, in
a $K$-causal spacetime a continuous $K^{+}$-utility satisfies $x<y$
$\Rightarrow (x,y)\in K^{+}\backslash\Delta \Rightarrow u(x)<u(y)$
and hence it is a time function. $\square$
\end{proof}

\begin{theorem} \label{vbn}
A spacetime is $K$-causal if and only if it  admits a time function
(as a consequence time functions are always $K^{+}$-utilities). In
this case, denoting with $\mathcal{A}$ the set of time functions
 we have that the partial order  $K^{+}$ can be recovered from the time functions,
that is
\begin{equation} \label{vgf}
(x,y) \in K^{+} \Leftrightarrow \forall t \in \mathcal{A}, \ t(x)\le
t(y).
\end{equation}
\end{theorem}

\begin{proof}
Assume that the spacetime admits a time function then it is
$K$-causal, that is $K^{+}$ is antisymmetric, indeed otherwise there
would be $p,q \in M$, $p\ne q$, such that $(p,q)\in K^{+}$ and
$(q,p)\in K^{+}$. By lemma \ref{jba}(b), $t(p)<t(q)<t(p)$, a
contradiction. By theorem \ref{ndc} we also infer that the time
function is a continuous $K^{+}$-utility.

Assume that the spacetime is $K$-causal then by Levin's theorem it
admits a continuous $K^{+}$-utility which by theorem \ref{ndc}  is a
time function.

The last statement is also an application of Levin's theorem.
$\square$
\end{proof}

Actually Levin's theorem states something more because it applies
also to the case in which $K$-causality does not hold. However, in
this case the $K^{+}$-utilities are not time functions.
Nevertheless, we have the following

\begin{lemma}
In a chronological spacetime in which $\overline{J^{+}}$ is
transitive (that is $K^{+}=\overline{J^{+}}$) the continuous
$K^{+}$-utilities are also continuous $I^{+}$-utilities, that is,
they are also semi-time functions.
\end{lemma}

\begin{proof}
Let $u$ be a $K^{+}$-utility, since the spacetime is chronological
we have only to prove $(x,y)\in I^{+} \Rightarrow u(x)<u(y)$. The
hypothesis is (i) $(x,y)\in \overline{J^{+}} \Rightarrow u(x)\le
u(y)$ and (ii) $(x,y)\in \overline{J^{+}}$ and $(y,x)\notin
\overline{J^{+}} \Rightarrow u(x)< u(y)$. Note that if $(x,y)\in
I^{+}$ then $(y,x) \notin \overline{J^{+}}$ because the relation
$I^{+}$ is open and the spacetime is chronological, thus $(x,y)\in
I^{+} \Rightarrow$ $(x,y)\in \overline{J^{+}}$ and $(y,x)\notin
\overline{J^{+}} \Rightarrow u(x)< u(y)$ which is the thesis.
$\square$
\end{proof}

As a consequence we are able to clarify that the consequences of
Levin's theorem are actually stronger than those of Peleg's theorem,
as we can now infer from theorem \ref{lev}

\begin{theorem}
A chronological spacetime in which $\overline{J^{+}}$ is transitive
 admits a semi-time function.
\end{theorem}

Of course it actually admits a continuous $K^{+}$-utility which is a
 stronger concept than that of semi-time function. We have expressed
 the theorem in this form for the sake of comparison with theorem \ref{weak}.

%
%Note that by lemma \ref{jba} point (a) a $K^{+}$-utility in the
%general case is any function $\tilde{t}$ such that
%\begin{equation}
%x\le y \to \Rightarrow \tilde{t}(x)\le \tilde{t}(y), \textrm{ and }
%\end{equation}

By using Levin's theorem and the smoothing result for time functions
\cite{bernal04} it is possible to give another proof of the
equivalence between $K$-causality and stable causality

\begin{theorem}
$K$-causality coincides with stable causality.
\end{theorem}

\begin{proof}
The proof that stable causality implies $K$-causality goes as usual.
The thesis follows from $K^{+}\subset J^{+}_S$, because $J^{+}_S$ is
closed, transitive and contains $J^{+}$ while $K^{+}$ is by
definition the smallest relation with this property. Thus this
direction follows from the equivalence between the antisymmetry of
$J^{+}_S$ and stable causality, the antisymmetry condition being
inherited by the inclusion of relations.

For the other direction $K$-causality implies the existence of a
time function, thus the existence of a temporal function and hence
stable causality. $\square$
\end{proof}

\section{Time orderings}
This section is independent of the previous one. Here the
representation theorem for $K^{+}$ (or $J^{+}_S$) through the time
functions is proved again without the help of Levin's theorem but
using the results of \cite{minguzzi08b}. I gave this proof  before
discovering the connection with utility theory. It is quite short
uses in an essential way the  equivalence between $K$-causality and
stable causality. In the last part of the proof I also use the
smoothability results of \cite{bernal04} in order to generalize to
temporal functions the representation theorem. This improvement is
important because it allows us to make a connection with `observers'
on spacetime provided we model them with congruences of timelike
curves.

%In this section we establish the following result which clarifies
%which aspects of the causal structure can be recovered starting from
%the time or temporal orderings allowed by the spacetime.

%It could be convenient to express Eq. (\ref{vgf}) in a different
%relational form.
% If the spacetime is $K$-causal there is at
%least one time function $t$, and to any such function there
%corresponds a total preorder of the spacetime events given by the
%relation

Given a time function on spacetime let us introduce the total
preorder
\begin{equation} \label{one}
T^{+}[t]=\{(p,q)\in M\times M: t(p)\le t(q) \}.
\end{equation}
Any such preorder, here called  {\em time ordering}, extends $J^{+}$
according to the definition of section \ref{secb}, in particular
$J^{+}\subset T^{+}[t]$. The relation $T^{+}[t]$ is closed because
$t$ is continuous. If $t$ is temporal then we shall say that the
time ordering $T[t]$ is also a {\em temporal ordering}.

Note that the relation $T^{+}[t]$ is invariant under monotonous time
reparametrizations, that is, if $f$ is increasing $f(t(\cdot))$ is a
time function and \[T^{+}[f(t)]=T^{+}[t].\] In other words, the
relation $T^{+}[t]$ keeps the information on the {\em simultaneity}
convention associated to the time function $t$, but it is
insensitive to the actual values of the time intervals $t(q)-t(p)$,
$p,q \in M$.

As a matter of convention, in the next intersections if the index
sets $\mathcal{A}$ or $\mathcal{B}$ are empty then the intersection
is the whole ambient space $M\times M$.

\begin{theorem} \label{bax}
In every spacetime
\[
K^{+}\subset J^{+}_S\subset \bigcap_{t\in \,\mathcal{A}}\, T^{+}[t]
\subset \bigcap_{t\in \,\mathcal{B}}\,  T^{+}[t].
\]
In a stably causal spacetime
\[
K^{+}= J^{+}_S= \bigcap_{t\in \,\mathcal{A}}\, T^{+}[t] =
\bigcap_{t\in \,\mathcal{B}}\,  T^{+}[t].
\]
\end{theorem}

\begin{figure}[ht]
\centering
%\psfrag{R}{{\small  Remove}} \psfrag{p}{{\small  p}}
%\psfrag{q}{{\small  q}} \psfrag{b}{{\small $\!\!t'=b$}}
%\psfrag{a}{{\small $t'=a$}} \psfrag{D}{{\small $D_N(S)$}}
%\psfrag{S}{{\small $S$}}
 \includegraphics[width=9cm]{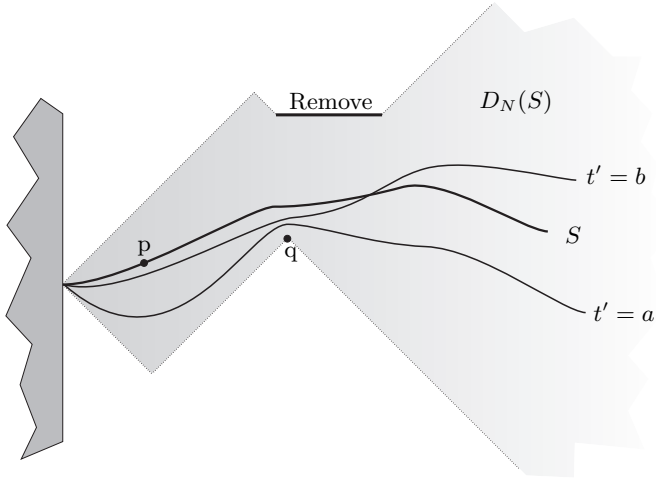}
 \caption{The last argument in the proof of theorem \ref{bax}.}
\label{pr}
\end{figure}

\begin{proof}
The first inclusion is well known while the latter is obvious
because $\mathcal{B}\subset \mathcal{A}$. If, by contradiction,
$J^{+}_S\subset \bigcap_{t\in \,\mathcal{A}}\, T^{+}[t]$ does not
hold then there is $(p,q)\in J^{+}_S\backslash \bigcap_{t\in
\,\mathcal{A}}\, T^{+}[t]$. In particular $\mathcal{A}$ is not empty
and there is a time function $t$ such that $t(p)>t(q)$. Now, note
that $J^{+}_S\cap \{\bigcap_{t\in \,\mathcal{A}}\, T^{+}[t]\}
\subsetneq J^{+}_S$ being the intersection of closed and transitive
relations which contain $J^{+}$, shares all these same properties.
As a consequence $K^{+}\ne J^{+}_S$, but it is known
\cite{minguzzi08b} that in a stably causal spacetime
$K^{+}=J^{+}_S$, thus the  spacetime is not stably causal although
there is a time function $t$, a contradiction.

Let $(M,g)$ be a stably causal spacetime. The equality
$K^{+}=J^{+}_S$ has been proved in \cite{minguzzi07,minguzzi08b}.
Let us prove $\bigcap_{t\in \,\mathcal{A}}\, T^{+}[t] \subset
J^+_S$. By contradiction, assume it does not hold, then there is a
pair $(p,q)\in \bigcap_{t\in \,\mathcal{A}}\, T^{+}[t] \backslash
J^{+}_S$. Recall \cite[Lemma 3.3]{minguzzi07} that
$J^{+}_S=\bigcap_{g'>g} \overline{J^{+}}_{g'}$. Since $(p,q) \notin
J^{+}_S$ there is $g'>g$ such that $p \notin \overline{J^-_{g'}(q)}$
and $(M,g')$ is  causal.

Let $A\ni p$ be an open
 set such that $A\cap
\overline{J^-_{g'}(q)}=\emptyset$. We are going to construct a time
function $\hat{t}$ such that $\hat{t}(p)>\hat{t}(q)$, a
contradiction with $(p,q)\in \bigcap_{t\in \,\mathcal{A}}\,
T^{+}[t]$. Basically we are going to use Hawking's averaging
technique \cite[Prop. 6.4.9]{hawking73}. We introduce a volume
measure $\mu$ as in \cite[Prop. 6.4.9]{hawking73} so that $\mu(M)$
is finite. We can find a family of Lorentz metrics $h(a)$, $a \in
[0,3]$, such that points (1)-(3) in that proof are satisfied and
$h(3)=g'$. Then we  construct a continuous function
$\bar{\theta}(x)=\int_1^2\theta(x,a) \dd a $ where
$\theta(x,a)=\mu(I^{-}_{h(a)}(x))$ as done there. However, here we
 make  just a little change. The measure $\mu$ is taken with support
in $A\cap I^{-}_g(p)$. As a consequence the function $\bar{\theta}$
is continuous and non-decreasing over every future directed causal
curve while in Hawking's construction it is increasing. Let $t$ be a
time function. The continuous function $\hat{t}=t+\bar{\theta}$ is a
time function and $\hat{t}(q)=t(q)$ while $\hat{t}(p)=t(p)+\mu(M)$.
By choosing $\mu(M)> t(q)-t(p)$ we get the thesis.

It remains to prove the inclusion $\bigcap_{t\in \,\mathcal{A}}
T^{+}[t] \supset \bigcap_{t\in \,\mathcal{B}} T^{+}[t]$. By
contradiction, suppose it does not hold then there is a pair $(p,q)
\in \bigcap_{t\in \,\mathcal{B}}  T^{+}[t]\backslash \bigcap_{t\in
\,\mathcal{A}} T^{+}[t]$. In other words for every temporal function
$t$ (there is at least one temporal function $\tau$ because $(M,g)$
is stably causal \cite{bernal04}), we have $t(p)\le t(q)$, but there
is a time function $\hat{t}$ such that $\hat{t}(p)>\hat{t}(q)$.
Consider the (acausal) partial Cauchy hypersurface
$S=\hat{t}^{-1}(\hat{t}(p))$, see figure \ref{pr}. The set $S$ does
not intersect $q$, let $N=M\backslash\{q\}$ so that $S$ is a partial
Cauchy hypersurface for $(N,g\vert_N)$. Let $D_N(S)$ be the Cauchy
development of $S$ on $(N,g\vert_N)$ and $H^{+}_N(S)$ and
$H^{-}_N(S)$ the future and past Cauchy horizons. We have $S\cap
H^{+}_N(S)=\emptyset$ because if $r \in S\cap H^{+}_N(S)$ then, as
$H^{+}_N(S)$ is generated by past inextendible lightlike geodesics
on $N$, there would be a past inextendible geodesic with future
endpoint $r$. No other point of the geodesic can belong to $S$
because of its acausality, but since $H^{+}_N(S)\backslash S\subset
I_N^{+}(S)$ we get a contradiction with the achronality of $S$.
Analogously, $S\cap H^{-}_N(S)=\emptyset$.

As a consequence the set $\textrm{Int} D_N(S)$ is non-empty and
being globally hyperbolic it is diffeomorphic to $\mathbb{R}\times
S$ where the slices diffeomeorphic to $S$ are the level sets of a
temporal function $t'$ on the spacetime $\textrm{Int} D_N(S)$ with
the induced metric (see \cite{bernal04}). Choosing  $a,b$, $a<b$, so
that $b<t'(p)$ and $a<b+\tau(p)-\tau(q)$, we construct a function
$t''$ on $\textrm{Int} D_N(S)$ so that $t''=t'$ at those points
where $a\le t'\le b$, $t''=b$ at those points such that $t'\ge b$,
and $t''=a$ at those points where $t'\le a$. Note that $t''(p)=b$.
Clearly $t''$ has past directed timelike gradient for $a<t''<b$ but
there is a discontinuity in the gradient for $t''=a$ or $t''=b$.
However, a smooth monotonous reparametrization $t'''=f(t'')$ exists
which sends $a$ to $a$, $b$ to $b$, and makes the gradient
everywhere continuous, timelike on $a<t'''<b$ and vanishing for
$t'''\le a$, and $t'''\ge b$. A possible choice is
\[
t'''=-\frac{b-a}{2}\cos [\pi \frac{t''-a}{b-a} ]+\frac{b+a}{2},
\quad \textrm{for } a\le t''\le b, \ t'''=t'' \textrm{ elsewhere}.
\]
The function $t'''$ can be extended in a smooth way to $M$ by
setting $t'''=b$ on $\hat{t}^{-1}((\hat{t}(p), +\infty))\backslash
\textrm{Int} D_N(S)$ and $t'''=a$ on $\hat{t}^{-1}((-\infty,
\hat{t}(p)))\backslash \textrm{Int} D_N(S)$. In particular, since $q
\notin \textrm{Int} D_N(S)$ and $\hat{t}(p)>\hat{t}(q)$ we have
$t'''(q)=a$.   The function $\check{t}=\tau+t'''$ is a temporal
function and $\check{t}(q)=\tau(q)+a<\tau(p)+b=\check{t}(p) $, a
contradiction. $\square$
\end{proof}

It must be remarked that to every temporal function $t$ there
corresponds a flow generated by the future directed timelike unit
vector $u=-\nabla t/\sqrt{-g(\nabla t,\nabla t)}$. The generated
congruence of timelike curves represents an extendend reference
frame so that every curve of the congruence is identified with an
observer ``at rest in the frame''. The flow is orthogonal to the
slices $t=const.$ which therefore are the natural simultaneity
slices as they would be obtained by the observers at rest in the
frame by a local application of the Einstein's simultaneity
convention \cite{minguzzi03,robb14,malament77}. This observation
shows that the temporal functions, at least in principle, can be
physically realized through a well defined operational procedure.
The above theorem then states that while observers living in
different extended reference frames may disagree on which event of a
pair comes ``before'' or ``after'' the other, according to their own
time function, they certainly agree whenever the pair of compared
events belong to the $K^{+}$ (Seifert) relation, and in fact only
for those type of pairs. In other words the $K^{+}$ (Seifert)
relation provides that set of pairs of events for which all the
observers agree on their temporal order.

%\begin{theorem}
%Let $(M,g)$ be a distinguishing spacetime. If there is a continuous
%function $\tilde{t}$ such that $x\le y \Rightarrow \tilde{t}(x)\le
%\tilde{t}(y)$, then $(M,g)$ is stably causal.
% \end{theorem}

Eq. (\ref{vgf}) can be rewritten in the equivalent form
\begin{equation}
K^{+}=\bigcap_{t\in \mathcal{A}} T^{+}[t],
\end{equation}
thus we have just obtained an alternative proof for the same
equation.

This result allows us to establish those circumstances in which the
chronological or causal relation can be recovered from the knowledge
of the time or temporal functions.

Recall that a spacetime is {\em causally easy} if it is strongly
causal and $\overline{J^{+}}$ is transitive \cite{minguzzi08b}.
Recall also that a causally continuous spacetime is a spacetime
which is distinguishing and reflective. Finally a spacetime is
causally simple \cite{bernal06b} if it is causal and
$\overline{J^{+}}=J^{+}$. We have causal simplicity $\Rightarrow$
causal continuity $\Rightarrow$ causal easiness $\Rightarrow$
$K$-causality.

By definition of causal easiness $K^{+}=\overline{J^{+}}$, thus as
$\overline{I^{+}}=\overline{J^{+}}$, we easily find

\begin{proposition} \label{ocf}
In a causally easy spacetime $I^{+}=\textrm{Int} \bigcap_t\,
T^{+}[t]$, and in a causally simple spacetime
$J^{+}=\bigcap_t\,T^{+}[t]$ where the intersections are with respect
the sets of time or the temporal functions.
\end{proposition}

\section{Conclusions}

The concept of causal influence is more primitive, and in fact more
intuitive, than that of time. General relativistic spacetimes have
by definition a causal structure but may lack a time function,
namely a continuous  function which respects the notion of causal
precedence (i.e. if $a$ influences $b$ then the time of $a$ is less
than that of $b$).

In this work we have recognized the mathematical coincidence between
the problem of the existence of a (semi-)time function on spacetime
in the relativistic physics field and the problem of the existence
of a utility function for an agent in microeconomics. From these
problems two so far independent lines of research arose which, as we
noted, often passed through the very same concepts. Remarkably, some
results obtained in one field were not rediscovered in the other, a
fact which has allowed us to use Peleg's and Levin's theorems to
reach new results concerning the existence of (semi-)time functions
in relativity.

In particular, we have  proved that a chronological spacetime  in
which $\overline{J^{+}}$  is transitive (for instance a  reflective
spacetime) admits a semi-time function. Also in a $K$-causal
spacetime the existence of a time function follows solely from the
closure and antisymmetry of the $K^{+}$ relation. In the other
direction we have proved without the help of smoothing techniques,
that the existence of a time function implies $K$-causality. We have
also given a new proof
 of the equivalence between $K$-causality and stable causality
by using Levin's theorem and smoothing techniques.

Finally, we have shown in two different ways that in a $K$-causal
(i.e. stably casual) spacetime the $K^{+}$ (i.e. Seifert) relation
can be recovered from the set of time or temporal functions allowed
by the spacetime. This result  singles out the $K^+$ relation as one
of the most important for the development of causality theory.

%\bibliography{../../bibliografie/simultaneity,../../bibliografie/libri,../../bibliografie/miei,../../bibliografie/mieiPreprints,../../bibliografie/mieiProceedings}

\begin{thebibliography}{10}

\bibitem{andrikopoulos09}
Andrikopoulos, A.: Szpilrajn-type theorems in economics (May 2009).
\newblock Mimeo, Univ. of Ionnina. Available at http://mpra.ub.unimuenchen.
  de/14345/

\bibitem{aumann62}
Aumann, R.~J.: Utility theory without the completeness axiom.
\newblock Econometrica \textbf{30}, 445--462 (1962)

\bibitem{beem76}
Beem, J.~K.: Conformal changes and geodesic completeness.
\newblock Commun. Math. Phys. \textbf{49}, 179--186 (1976)

\bibitem{bernal04}
Bernal, A.~N. and {S\'a}nchez, M.: Smoothness of time functions and
the metric
  splitting of globally hyperbolic spacetimes.
\newblock Commun. Math. Phys. \textbf{257}, 43--50 (2005)

\bibitem{bernal06b}
Bernal, A.~N. and {S\'a}nchez, M.: Globally hyperbolic spacetimes
can be
  defined as {`causal'} instead of {`strongly causal'}.
\newblock Class. Quantum Grav. \textbf{24}, 745--749 (2007)

\bibitem{bossert99}
Bossert, W.: Intersection quasi-orderings: An alternative proof.
\newblock Order \textbf{16}, 221--225 (1999)

\bibitem{bridges95}
Bridges, D.~S. and Mehta, G.~B.: \emph{Representations of preference
  orderings}, vol. 442 of \emph{Lectures Notes in Economics and Mathematical
  Systems}.
\newblock Berlin: {Springer-Verlag} (1995)

\bibitem{candeal93}
{Candeal-Haro}, J.~C. and {Indur\'ain-Eraso}, E.: Utility
representations from
  the concept of measure.
\newblock Mathematical {S}ocial {S}ciences \textbf{26}, 51--62 (1993)

\bibitem{clarke88}
Clarke, C. J.~S. and Joshi, P.~S.: On reflecting spacetimes.
\newblock Class. Quantum Grav. \textbf{5}, 19--25 (1988)

\bibitem{debreu54}
Debreu, G.: \emph{Representation of preference ordering by a
numerical
  function}, New York: John Wiley, vol. Decision Processes, ed. R. M. Thrall
  and C. H. Coombs and R. L. Davis, pages 159--165 (1954)

\bibitem{debreu64}
Debreu, G.: Continuity properties of {P}aretian utility.
\newblock International {E}conomic {R}eview \textbf{5}, 285--293 (1964)

\bibitem{dieckmann88}
Dieckmann, J.: Volume functions in general relativity.
\newblock Gen. Relativ. Gravit. \textbf{20}, 859--867 (1988)

\bibitem{donaldson98}
Donaldson, D. and Weymark, J.~A.: A quasiordering is the
intersection of
  orderings.
\newblock Journal of {E}conomic {T}heory \textbf{78}, 328--387 (1998)

\bibitem{dushnik41}
Dushnik, B. and Miller, E.: Partially ordered sets.
\newblock Amer. J. Math. \textbf{63}, 600--610 (1941)

\bibitem{eilenberg41}
Eilenberg, S.: Ordered topological spaces.
\newblock American {J}ournal of {M}athematics \textbf{63}, 39--45 (1941)

\bibitem{evren09}
Evren, O. and Ok, E.~A.: On the multi-utility representation of
preference
  relations (June 2008)

\bibitem{geroch70}
Geroch, R.: Domain of dependence.
\newblock J. Math. Phys. \textbf{11}, 437--449 (1970)

\bibitem{hawking68}
Hawking, S.~W.: The existence of cosmic time functions.
\newblock Proc. {R}oy. {S}oc. {L}ondon, series {A} \textbf{308}, 433--435
  (1968)

\bibitem{hawking73}
Hawking, S.~W. and Ellis, G. F.~R.: \emph{The Large Scale Structure
of
  Space-Time}.
\newblock Cambridge: Cambridge {U}niversity {P}ress (1973)

\bibitem{hawking74}
Hawking, S.~W. and Sachs, R.~K.: Causally continuous spacetimes.
\newblock Commun. Math. Phys. \textbf{35}, 287--296 (1974)

\bibitem{herden89}
Herden, G.: On the existence of utility functions.
\newblock Mathematical {S}ocial {S}ciences \textbf{17}, 297--313 (1989)

\bibitem{herden95}
Herden, G.: On some equivalent approaches to mathematical utility
theory.
\newblock Mathematical {S}ocial {S}ciences \textbf{29}, 19--31 (1995)

\bibitem{kim93}
Kim, J.-C. and Kim, J.-H.: Totally vicious spacetimes.
\newblock J. Math. Phys. \textbf{34}, 2435--2439 (1993)

\bibitem{lee72}
Lee, L.-F.: The theorems of {D}ebreu and {P}eleg for ordered
topological
  spaces.
\newblock Econometrica \textbf{40}, 1151--1153 (1972)

\bibitem{levin83}
Levin, V.~L.: A continuous utility theorem for closed preorders on a
  $\sigma$-compact metrizable space.
\newblock Soviet Math. Dokl. \textbf{28}, 715--718 (1983)

\bibitem{malament77}
Malament, D.~B.: Causal theories of time and the conventionality of
  simultaneity.
\newblock No\^us \textbf{11}, 293--300 (1977)

\bibitem{mehta77}
Mehta, G.: Topological ordered spaces and utility functions.
\newblock International {E}conomic {R}eview \textbf{18}, 779--782 (1977)

\bibitem{mehta83}
Mehta, G.: Ordered topological spaces and the theorems of {D}ebreu
and {P}eleg.
\newblock Indian {J}. {P}ure {A}ppl. {M}ath. \textbf{14}, 1174--1182 (1983)

\bibitem{minguzzi03}
Minguzzi, E.: Simultaneity and generalized connections in general
relativity.
\newblock Class. Quantum Grav. \textbf{20}, 2443--2456 (2003)

\bibitem{minguzzi07b}
Minguzzi, E.: The causal ladder and the strength of {$K$}-causality.
{I}.
\newblock Class. Quantum Grav. \textbf{25}, 015009 (2008)

\bibitem{minguzzi07}
Minguzzi, E.: The causal ladder and the strength of {$K$}-causality.
{II}.
\newblock Class. Quantum Grav. \textbf{25}, 015010 (2008)

\bibitem{minguzzi07c}
Minguzzi, E.: Limit curve theorems in {L}orentzian geometry.
\newblock J. Math. Phys. \textbf{49}, 092501 (2008)

\bibitem{minguzzi07f}
Minguzzi, E.: Non-imprisonment conditions on spacetime.
\newblock J. Math. Phys. \textbf{49}, 062503 (2008)

\bibitem{minguzzi08b}
Minguzzi, E.: {$K$}-causality coincides with stable causality.
\newblock {C}ommun. {M}ath. {P}hys. \textbf{290}, 239--248 (2009)

\bibitem{minguzzi06c}
Minguzzi, E. and S\'anchez, M.: \emph{The causal hierarchy of
spacetimes},
  Zurich: Eur. Math. Soc. Publ. House, vol. H. Baum, D. Alekseevsky (eds.),
  Recent developments in pseudo-Riemannian geometry, of \emph{{ESI} Lect. Math.
  {P}hys.}, pages 299--358 (2008).
\newblock ArXiv:gr-qc/0609119

\bibitem{nachbin65}
Nachbin, L.: \emph{Topology and order}.
\newblock Princeton: D. {V}an {N}ostrand {C}ompany, {I}nc. (1965)

\bibitem{nomizu61}
Nomizu, K. and Ozeki, H.: The existence of complete riemannian
metrics.
\newblock Proc. {A}mer. {M}ath. {S}oc. \textbf{12}, 889--891 (1961)

\bibitem{peleg70}
Peleg, B.: Utility functions for partially ordered topological
spaces.
\newblock Econometrica \textbf{38}, 93--96 (1970)

\bibitem{rader63}
Rader, T.: The existence of a utility function to represent
preferences.
\newblock The {R}eview of {E}conomic {S}tudies \textbf{30}, 229--232 (1963)

\bibitem{robb14}
Robb, A.~A.: \emph{A Theory of Time and Space}.
\newblock Cambridge: Cambridge {U}niversity {P}ress (1914)

\bibitem{seifert71}
Seifert, H.: The causal boundary of space-times.
\newblock Gen. Relativ. Gravit. \textbf{1}, 247--259 (1971)

\bibitem{seifert77}
Seifert, H.~J.: Smoothing and extending cosmic time functions.
\newblock Gen. Relativ. Gravit. \textbf{8}, 815--831 (1977)

\bibitem{sondermann80}
Sondermann, D.: Utility representations for partial orders.
\newblock Journal of {E}conomic {T}heory \textbf{23}, 183--188 (1980)

\bibitem{sorkin96}
Sorkin, R.~D. and Woolgar, E.: A causal order for spacetimes with
{$C^0$}
  {L}orentzian metrics: proof of compactness of the space of causal curves.
\newblock Class. Quantum Grav. \textbf{13}, 1971--1993 (1996)

\bibitem{szpilrajn30}
Szpilrajn, E.: Sur l'extension de l'ordre partiel.
\newblock Fund. Math. \textbf{16}, 386--389 (1930)

\bibitem{ward54}
{Ward, Jr.}, L.~E.: Partially ordered topological spaces.
\newblock Proc. Am. Math. Soc. \textbf{5}, 144--161 (1954)

\end{thebibliography}
%\bibliographystyle{cmp}
%\bibliographystyle{plain}

\section*{Acknowledgments}
This work has been partially supported by GNFM of INDAM and by FQXi.

\end{document}